\documentclass[prl,twocolumn,floatfix,amsfonts]{revtex4}
\usepackage{graphicx,graphics,color,epsfig}
\usepackage{bm}
\usepackage{amsmath}
\usepackage{amssymb}
\begin{document}
\title{Reply to a Comment of M. Continentino on
``Universally diverging Gr\"uneisen parameter and the magnetocaloric effect
close to quantum critical points''}
\author{Lijun Zhu$^1$, Markus Garst$^2$, Achim Rosch$^3$, and Qimiao Si$^1$}
\address{$^1$Department of Physics \& Astronomy, Rice University, Houston,
TX 77005--1892, USA\\
$^2$Institut f\"ur Theorie der Kondensierten Materie, Universit\"at Karlsruhe,
D-76128 Karlsruhe, Germany\\
$^3$ Institut f\"ur Theoretische Physik, University of Cologne,
50937 K\"oln, Germany}

\begin{abstract}
We show that the comment [cond-mat/0408217] by Continentino on our 
recent paper [PRL {\bf 91}, 066404 (2003)] reaches incorrect conclusions
as the comment wrongly extrapolates from results valid  close to a classical 
phase transition into the quantum critical regime.
\end{abstract}

\maketitle

In his comment \cite{continentino} and a previous preprint \cite{cont2}
Continentino claims that our result~\cite{zhu}
for the quantum-critical Gr\"uneisen ratio
$\Gamma=\frac{\alpha}{c_p} \sim T^{-1/(\nu z)}$ 
(where $\alpha$ and $c_p$ are thermal expansion and
specific heat, respectively)
should be replaced by $\Gamma \sim
T^{-1/\psi}$, where $\psi$ is the shift exponent.
We explain below why this claim is incorrect
above the upper critical dimension, {\it i.e.} $d+z > 4$.
Below the upper critical dimension, one has $\psi = \nu z$
and no discrepancy arises.

We will focus on the {\it quantum critical regime}, about which
the comment raises the issue; the experiments reported in
Ref.~\cite{exp} also concern this regime.
Above the upper critical dimension ($d+z>4$),
a scaling
analysis (see below) is more subtle due to the presence of some dangerously
irrelevant variable(s), but a direct calculation of $\Gamma$ 
in the quantum critical regime is straightforward.
Therefore we reported in \cite{zhu} the results of explicit calculations
for spin-density-wave (SDW) quantum critical 
points (QCPs) \cite{Hertz,Millis,Sachdev},
where the quartic
coupling is dangerously irrelevant
and $\psi \ne \nu z$. 
For instance, for $d=3$ and $z=2$ where the exponents
$\psi = 2/3$ and $\nu z=1$, we find 
$\Gamma \sim T^{-1}$ while Continentino would predict
$\Gamma \sim T^{-3/2}$. These calculations already
show that the claim of Continentino is incorrect.


\begin{figure}[h]
\begin{center}
\includegraphics[width=0.7\linewidth]{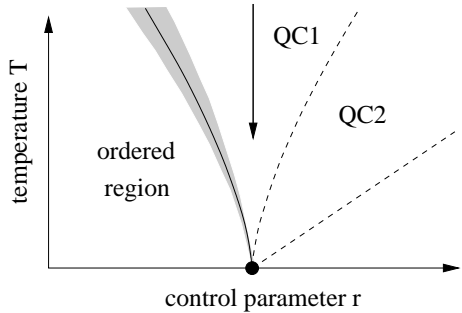}
\end{center}
\caption{%
Schematic phase diagram near a quantum-critical point.
In the shaded area, classical scaling is expected.
The temperature dependence in the quantum critical regime is
measured along a path indicated by the arrow.
The dashed lines, defined for positive $r$,
describe the upper and lower crossover temperatures,
$T_{\rm cr1} \propto (r/u)^{\psi} $
and $T_{\rm cr2} \propto r^{\nu z}$,
respectively.
}
\label{phasedia}
\end{figure}
 It is straightforward to pin down why the rather general argument given
 in the comment fails. The reason is that both
  the Ehrenfest relation \cite{continentino} and the scaling ansatz 
  used in Ref.~\cite{cont2} [Eq.~(5) of Ref.~\cite{cont2}; the ansatz is
a special form of Eq. (1.30) in Ref.~\cite{Continentino-book}]
are only valid close
 to the {\em classical} phase transition, i.e. in the shaded area shown in
  Fig.~\ref{phasedia}, but not in the quantum critical region.
  The part of the calculation of Ref.~\cite{zhu} under discussion is,
  however, performed along the arrow depicted in Fig.~\ref{phasedia}
  which never enters the shaded region.

  The problem with Continentino's approach can be traced back to
the qualitatively different role which the quartic coupling $u>0$
of magnetic fluctuations plays
 in the classical critical and quantum
  critical regimes: it is a relevant coupling in the former but
  (dangerously) irrelevant in the latter.  This  makes it
  incorrect to use the simple scaling form Eq. (5) of Ref.~\cite{cont2} to
  describe all the regimes. Instead, one has to include
explicitly the quartic coupling $u$
in the scaling ansatz for the free energy
\begin{eqnarray}
F\approx T^{\frac{d+z}{z}} f\!\left(\frac{r}{T^{1/(\nu z)}},
u T^{\frac{d+z-4}{z}}\right)
\end{eqnarray}
where $r \propto p-p_c$ measures the distance from the QCP. We assume
$d+z>4$, $d>1/\nu$ and $d\neq z$ to avoid logarithmic terms. In the
quantum critical regimes QC1 and QC2,
$|x|,y\ll 1$ and $x+y \gg y^\frac{1}{\nu (4-d)}$,
one obtains (omitting all
multiplying constant factors)
$f(x,y)\approx 1+ x+ y$.
($x$ and $y$ appear separately in the Fermi liquid regime below QC2.)
In contrast, the scaling ansatz \cite{cont2,Continentino-book} 
of Continentino
$f \propto
|x+y|^{2-\tilde{\alpha}}$, can only be valid within the Ginzburg
region of the classical transition (shaded area in Fig.~\ref{phasedia}),
i.e. only for $|x+y| \ll y^\frac{1}{\nu (4-d)}$ for $d<4$
(see, {\it e.g.}, Ref.~\cite{Millis}).
The use of this formula outside of its range of applicability in region QC1
(as done in Refs.~\cite{cont2,Continentino-book} and implicitly 
when comparing
 our results to the Ehrenfest relation in Ref.~\cite{continentino})
 implies that $F \propto u^{2-\tilde{\alpha}}
T^{\frac{\tilde{\alpha}-\alpha}{\nu z}+ \frac{2
    -\tilde{\alpha}}{\psi}}$ for $r=0$. This result is wrong on two
accounts. First, it entirely misses the leading term which
corresponds to Gaussian critical fluctuations
[cf.~the ``1'' term in the expression for $f(x,y)$ above].
Second, it is incorrect even for the sub-leading term
for $\psi \neq \nu z$ as it depends explicitly on the exponent
$\tilde{\alpha}$ which characterizes only the {\em classical} but not
the quantum-critical transition (in some cases, e.g. for $d=2$ and
Heisenberg symmetry, a classical transition does not even exist).
It therefore also contradicts numerous results in the literature, e.g.
Refs.~\cite{Millis,Sachdev}.

In his comment \cite{continentino} Continentino stated that Gaussian 
fluctuations should
not be taken into account in the scaling analysis.
  This is incorrect, since  Gaussian fluctuations are part of the
  singular contributions
at the QCP and in fact
  the leading singular terms in the QC1 and QC2 regions.
  As explained above, the dominance of the Gaussian  fluctuations in QC1 
and QC2
  is fully compatible (see e.g. Refs.~\cite{Millis,Sachdev}) with both 
the existence of a N\'eel line and $\psi \ne \nu z$ due to the dangerously
irrelevant variable $u$,.

  To summarize, the comment \cite{continentino} reaches incorrect 
conclusions by extrapolating from results valid very close to 
the classical phase transition into the quantum critical regime 
where they are not valid. Furthermore the scaling ansatz used in
Refs.~\cite{cont2,Continentino-book} incorrectly treats the dangerously
irrelevant quartic coupling
above the upper critical dimension.



\begin{thebibliography}{99}

\bibitem{continentino} M. Continentino,
cond-mat/0408217v1.
\bibitem{zhu} L.~Zhu,
M.~Garst, A.~Rosch, and Q.~Si,
Phys.\ Rev.\ Lett.\ {\bf 91}, 066404 (2003).
\bibitem{cont2} M. Continentino, cond-mat/0307218.
\bibitem{exp}
R.~K\"uchler {\it et al.}, Phys.\ Rev.\ Lett.\ {\bf 91}, 066405 (2003).
\bibitem{Continentino-book} M. A. Continentino,
{\it Quantum Scaling in Many Body Systems},
(World Scientific, Singapore, 2001), Chap.~1.
\bibitem{Hertz}
J.\ A.\ Hertz, 
Phys.\ Rev.\ B {\bf 14}, 1165 (1976).
\bibitem{Millis} A.\ J.\ Millis,
Phys.\ Rev.\ B {\bf 48}, 7183 (1993).
\bibitem{Sachdev} S. Sachdev, {\it Quantum Phase Transitions},
Cambridge University Press, Cambridge (1999).

\end{thebibliography}
\end{document}